\documentclass[aip, chaos,
amsmath,amssymb,
 reprint,%
]{revtex4-1}

\usepackage{graphicx}% Include figure files
\usepackage{dcolumn}% Align table columns on decimal point
\usepackage{bm}% bold math
\usepackage[utf8]{inputenc}
\usepackage[T1]{fontenc}
\usepackage{mathptmx}
\usepackage{etoolbox}

\usepackage{amssymb}
\usepackage{amsmath}
\usepackage{subcaption}
\usepackage{graphicx}
\usepackage{epstopdf}
\usepackage{physics}
\usepackage{soul}
\usepackage[dvipsnames]{xcolor}
\usepackage{booktabs}
\usepackage{scalerel}
\usepackage{lineno}
\usepackage{physics}
\newcommand{\diff}{\mathrm{d}} % Defines a command for an upright differential 'd'

\usepackage{hyperref}

\makeatletter
\def\@email#1#2{%
 \endgroup
 \patchcmd{\titleblock@produce}
  {\frontmatter@RRAPformat}
  {\frontmatter@RRAPformat{\produce@RRAP{*#1\href{mailto:#2}{#2}}}\frontmatter@RRAPformat}
  {}{}
}%
\makeatother
\begin{document}

\preprint{AIP/123-QED}

\title[]{Collective vibrational resonance and mode selection in nonlinear resonator arrays}
% Force line breaks with \\
\author{Somnath Roy}
 \affiliation{Institute of Engineering \& Management, University of Engineering \& Management,\\Kolkata,700091, West Bengal,India}
 \email{roysomnath63@gmail.com}

\author{Mattia Coccolo}%
\email{mattiatommaso.coccolo@urjc.es}
\affiliation{Nonlinear Dynamics, Chaos and Complex Systems Group,\\ 
Departamento de Física, Universidad Rey Juan Carlos,\\
Tulipán
s/n, 28933 Móstoles, Madrid, Spain}%

\author{Anirban Ray}%
\email{anirban.chaos@gmail.com}
\affiliation{Department of Physics,\\
Gour Mahavidyalaya, Mangalbari,\\
Malda 732142, India}%

\author{Asesh Roy Chowdhury}
\email{arc.roy@gmail.com}
\affiliation{Jadavpur University, Kolkata, 700032, West Bengal, India%\\This line break forced% with \\
}%

\begin{abstract}
This article investigates how a uniform high frequency (HF) drive applied to each site of a weakly-coupled discrete nonlinear resonator array can modulate the onsite natural stiffness and damping and thereby facilitate the active tunability of the nonlinear response and the phonon dispersion relation externally. Starting from a canonical model of parametrically excited \textit{van der Pol-Duffing} chain of oscillators with nearest neighbor coupling, a systematic two-widely separated time scale expansion (\textit{Direct Partition of Motion}) has been employed, in the backdrop of Blekhman's perturbation scheme. This procedure eliminates the fast scale and yields the effective collective dynamics of the array with renormalized stiffness and damping, modified by the high-frequency drive. The resulting dispersion shift controls which normal modes enter the parametric resonance window, allowing highly selective activation of specific bulk modes through external HF tuning. The collective resonant response to the parametric excitation and mode-selection by the HF drive has been analyzed and validated by detailed numerical simulations. The results offer a straightforward, experimentally tractable route to active control of response and channelize energy through selective mode activation in MEMS/NEMS arrays and related resonator platforms.
\end{abstract}

\maketitle

\begin{quotation}
Regulating the collective behavior of coupled nonlinear oscillators is the backbone of the development of emerging phononic and MEMS/NEMS technologies. Although conventional methods are limited by the static fabrication procedure, a dynamic approach to actively adjust the spectral response of a discrete nonlinear array is investigated in this work. The application of a consistent and uniform HF drive to each element can effectively modify the onsite stiffness and damping of the chain via a nonlinear averaging process. Through a systematic averaging and multiple-time-scale perturbation analysis, closed-form analytical results are obtained, connecting the external drive parameters to the internal dispersion properties. These theoretical predictions are thoroughly confirmed via numerical simulations, validating that the HF drive plays as an exact control mechanism to firmly adjust the band shape. This phenomenon facilitates the accurate excitation of specific bulk propagating modes, which allows the system to switch between various spatial patterns simply by adjusting the HF drive amplitude. The presented results connect dynamical theory with practical device engineering, providing an experimentally amenable framework for advancing programmable phononic crystals and adaptive sensors without requiring intricate, site-specific alteration.
\end{quotation}
\section{\label{sec:level1}Introduction}

Over the last few decades, advancements in the field of materials science have witnessed a paradigm shift from naturally occurring substances to artificially engineered media with altered wave propagation characteristics. The periodic structure of materials that are tailored to manipulate the mechanical waves is the leading area of studies in recent years, in which phononic crystals and acoustic meta-materials have captured a significant role \cite{fang2024advances,muhammad2022photonic,liu2020review,wang2020tunable}. It is possible to create materials that can be designed meticulously by using repeating unit cells. These artificially fabricated materials exhibit phenomena like vibration isolation by changing the energy band gap, mitigating acoustic energy by guided waves, and signal processing by spatial filtering. These features of phononic crystals and meta-materials have well-accounted applications from noise cancellation and seismic wave protection to advanced imaging, sensitive and tunable sensors, and thermal management \cite{olsson2008microfabricated,pennec2010two,dalela2022review,chen2012metamaterials}.\\

\begin{figure*}[t]
\centering
\includegraphics[width=\textwidth]{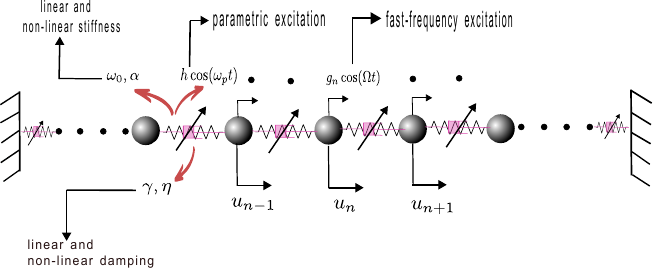}
    \caption{A schematic model of a coupled parametrically excited van der Pol-Duffing resonator array with fixed ends. Each site is excited by an additive high-frequency drive $g_n\cos(\Omega t)$.}
    \label{fig:schematic}
\end{figure*}

A powerful paradigm emerges as dynamic tuning over the static and fixed properties of the materials during fabrication. The active tuning under the effect of an external stimuli, can effectively alter the characteristics of the materials in real-time. This perspective transforms a static characteristic into an adaptive one, leading to a conceptual jump from ``designed matter'' to ``programmable matter''. From this perspective of a controlled mechanism, the arrays of MEMS (micro electro-mechanical systems) or NEMS (nano electro-mechanical systems) serve as an experimental platform for realizing those crystal structures and provide the physical intuition about the collective behavior of the lattice \cite{buks2002electrically}. Several studies pertaining to the dynamic tuning of these arrays facilitate the transformation of the materials into a programmable system, which is necessary for next-generation engineering devices \cite{lifshitz2008nonlinear,wei2021recent,kharrat2010modal}. Finally, the relationship is bidirectional: the knowledge about the engineered materials has also been employed to design and enhance the performance of MEMS/NEMS devices themselves \cite{mohammadi2012waveguide,choi2021fabricating}.\\

In this context, arrays of coupled van der Pol oscillators provide a canonical model for understanding the workings of mechanically coupled resonators. The van der Pol arrays, along with different modes of excitations, have been studied extensively to capture the interplay between the intrinsic complex nonlinearity and the external modulation \cite{lifshitz2003response,pinto2002collective,bitar2015collective,enjieu2007spatiotemporal,kashchenko2021dynamics}. The pivotal aspects of considering this model reside in its ability to explain a plethora of collective phenomena, from synchronization in discrete classical and quantum systems \cite{cross2004synchronization,wachtler2023topological,wolfovich2025modal} to intrinsic localized modes in continuous media \cite{kenig2009intrinsic}. These characteristics are comprehended from their applications ranging from neural network \cite{shuai2022synchronization}, power grids \cite{johnson2015synthesizing,sinha2015uncovering}, to neuromorphic computing \cite{shougat2023van,raychowdhury2018computing} and MEMS/NEMS devices \cite{hoppensteadt2002synchronization,asadi2018nonlinear,fon2017complex,aubin2004limit}. Although the studies pertaining to the behavior of resonator arrays subjected to external excitation are well established in the literature \cite{lifshitz2003response,bitar2015collective}, this article investigates a potentially tractable method for the dynamic control of collective response in such systems under the effect of a high-frequency (HF) force, applied uniformly to each site of the array. This well-acknowledged phenomena, termed as \textit{Vibrational Resonance} (VR) \cite{landa2000vibrational}, has been reported in a wide range of applications in the field of biology, engineering, and physical sciences \cite{yang2024vibrational}. Among the extensive body of work, several pivotal studies have been investigated in diverse contexts, including different nonlinear potentials \cite{rajasekar2011novel,ullner2003vibrational}, signal detection \cite{chizhevsky2008vibrational}, energy harvesting \cite{coccolo2014energy}, and quantum system \cite{olusola2020quantum,roy2025controlling}. While applications of VR have been corroborated in different studies concerning a single parametrically excited van der Pol oscillator \cite{roy2021vibrational,fahsi2009effect,belhaq20082} and also in neural systems \cite{deng2010vibrational,yu2011vibrational}, its systematic application and analysis to array oscillators, despite its relevance to diverse fields as discussed, to the best of our knowledge, has not been reported in the literature. This article addresses this gap by investigating the vibrational signal-driven response of self-sustained chain oscillators. The investigation also demonstrates how fundamental collective properties of the system can be tuned effectively by considering the average effect of the signal. In particular, it has been shown that the effective onsite stiffness and damping are reshaped by the application of a uniform high-frequency (HF) additive drive by intricate conjugation of the nonlinearities present in the system, thereby modulating the dispersion landscape of the array structure. This tunable method by HF drive for creative collective oscillatory patterns is made apparent by the modulated dispersion shift, which empowers the activation of a selective mode that is responsible for \emph{bulk mode wave} propagation throughout the array. The direction is important for phononic and MEMS/NEMS platforms because it offers a straightforward and experimentally feasible method of achieving real-time spectral control and programmable mode selection without requiring complicated spatial modulation or multi-site actuation.

\section{Model}

We consider a one-dimensional array of $N$ weakly nonlinear self-sustained oscillators with nearest-neighbor coupling, subject to a uniform high-frequency additive forcing. The motivation behind the adoption of this model relies on the foundational work of Lifshitz and Cross \cite{lifshitz2003response}, which introduced a minimal theoretical framework for the explanation of an experiment involving a parametrically excited MEMS resonator array \cite{buks2002electrically}
\begin{equation}
\begin{split}
\ddot{u}_{n}&+\epsilon\gamma\dot{u}_{n}+(\omega_{o}^{2}+\epsilon h \cos \omega_{p}t)u_{n}\\
&+\frac{1}{2}\epsilon d(u_{n+1}-2u_{n}+u_{n-1}) + \epsilon\alpha u_{n}^{3} + \epsilon\eta u_{n}^{2}\dot{u}_{n} = g_{n}\cos\Omega t
\end{split}
\label{eq:EOM}
\end{equation}
for \(n=1,\dots, N\), with fixed boundary conditions $$
u_{0} = u_{N+1} = 0
$$
Here, \(u_n(t)\) denotes the displacement of the \(n\)-th oscillator, 
\(\omega_0\) is the natural frequency of the particular site. Also, each resonator is parametrically excited by the drive $h\cos(\omega_p t)$. The term \(d\) represents the nearest-neighbor coupling strength. The positive $d$ is for a restoring interaction. The coefficient \(\alpha\) characterizes the cubic nonlinear stiffness (Duffing) and is taken as positive for hardening response. \(\gamma\) is the linear dissipation constant with an amplitude-dependent nonlinear damping contribution represented by the parameter \(\eta\). In addition, the key modification that is incorporated by applying a uniform high-frequency \(g_n \cos(\Omega t)\) additive forcing to each site, where $g_n$ specifies the forcing amplitude and \(\Omega\) is the fast frequency. It is assumed that $\Omega$ is \emph{nonresonant} and \emph{large} compared to all the characteristic frequencies ($\omega_0, \omega_p$) of the system. Also \(\Omega \gg \omega_{\max}\), where \(\omega_{\max}\) is the highest linear lattice eigen frequency (\(\approx\sqrt{\omega_0^2+4d}\)). The relative order of $g_n\sim\mathcal{O}(\Omega^2)$, such that $\dfrac{g_n}{\Omega^2}\sim\mathcal{O}(1)$. Under this scaling, all perturbative contributions $\left(\gamma,h,\alpha,\eta,d\right)$$\sim\mathcal{O}(1)$, enter at the same order $\epsilon$, allowing the system parameters to be treated on equal footing within the asymptotic expansion. The analytical procedure to investigate Eq.\eqref{eq:EOM} serves a two-fold purpose: First, to tackle the fast forcing, the \emph{Blekhman's} perturbation, also known as \emph{direct partition motion}, has been applied to average out the fast scale, yielding the effective dynamics. Second, the multiple time scale is incorporated to analyze the effective system in order to obtain the collective amplitude flow equations, and consequently, the modified dispersion relation.\\

By introducing a partition of time scales, $u_n$ can be decomposed into a slowly varying component $U_n$ and a fast oscillating component $\Psi_n$:
\begin{equation}
u_{n} = U_{n} (\omega_0 t,\omega_p t;t) + \Psi_{n}(\Omega t;t)
\label{eq:scale_separation}
\end{equation}
Substituting this into the governing equation (1) yields:
\begin{equation}
\begin{split}
    (\ddot{U}_{n}+\ddot{\Psi}_{n}) &+ \epsilon\gamma(\dot{U}_{n}+\dot{\Psi}_{n}) + (\omega_{o}^{2}+\epsilon h \cos \omega_{p}t)(U_{n}+\Psi_{n}) \\
    &+ \frac{1}{2}\epsilon d(U_{n+1}+\Psi_{n+1}-2(U_{n}+\Psi_{n})+U_{n-1}+\Psi_{n-1}) \\
    &+ \epsilon\alpha(U_{n}^{3}+3U_{n}^{2}\Psi_{n}+3U_{n}\Psi_{n}^{2}+\Psi_{n}^{3}) \\
    &+ \epsilon\eta(U_{n}^{2}+2U_{n}\Psi_{n}+\Psi_{n}^{2})(\dot{U}_{n}+\dot{\Psi}_{n}) = g_{n}\cos\Omega t
    \label{eq:slow_fast_EOM}
    \end{split}
\end{equation}
where \(U_n\) varies on time scales comparable to $\mathcal{O}(1/\omega_0)$ or slower, while the time scale of \(\Psi_n\) contains the fast oscillation at \(\Omega\). The slow component \(U_n\) is treated as quasi-constant when solving for \(\Psi_n\) by taking the average of Eq.~\eqref{eq:slow_fast_EOM} on the fast time scale, noting that the average of the fast variable and its derivatives is zero, while the slow variable remains unchanged:
$$
\langle\Psi_{n}\rangle = \langle\dot{\Psi}_{n}\rangle = \langle\ddot{\Psi}_{n}\rangle = \dots = 0
$$
$$
\langle u_{n}\rangle = \frac{1}{T}\int_{0}^{T} u_{n} dt = U_{n}
$$
The averaged equation for the slow dynamics becomes:
\begin{equation}
\begin{split}
    \ddot{U}_{n}&+\epsilon\gamma\dot{U}_{n}+(\omega_{o}^{2}+\epsilon h \cos \omega_{p}t)U_{n}+\frac{1}{2}\epsilon d(U_{n+1}-2U_{n}+U_{n-1}) \\
    &+ \epsilon\alpha(U_{n}^{3}+3U_{n}\langle\Psi_{n}^{2}\rangle+\langle\Psi_{n}^{3}\rangle) + \epsilon\eta(U_{n}^{2}+\langle\Psi_{n}^{2}\rangle)\dot{U}_{n} \\
    &+ \epsilon\eta(2U_{n}\langle\Psi_{n}\dot{\Psi}_{n}\rangle+\langle\Psi_{n}^{2}\dot{\Psi}_{n}\rangle) = \langle g_{n}\cos\Omega t\rangle \quad 
    \end{split}
    \label{eq:effective_average_slow}
\end{equation}

Subtracting the slow dynamics equation Eq.~\eqref{eq:effective_average_slow}  from the equation Eq.~\eqref{eq:slow_fast_EOM} gives the dynamics for the fast variable $\Psi_n$:
\begin{equation}
\begin{split}
    \ddot{\Psi}_{n} &+ \epsilon\gamma\dot{\Psi}_{n} + (\omega_{o}^{2}+\epsilon h \cos \omega_{p}t)\Psi_{n}
    + \frac{1}{2}\epsilon d(\Psi_{n+1}-2\Psi_{n}+\Psi_{n-1}) \\
    &+ \epsilon\alpha(3U_{n}^{2}\Psi_{n} + 3U_{n}(\Psi_{n}^{2} - \langle\Psi_{n}^{2}\rangle) + \dots) \\
    &+ \epsilon\eta(2U_{n}\Psi_{n}+\Psi_{n}^{2}-\langle\Psi_{n}^{2}\rangle)\dot{U}_{n}\\ &+ \epsilon\eta(U_{n}^{2}\dot{\Psi}_{n}+2U_{n}(\Psi_{n}\dot{\Psi}_{n}-\langle\Psi_{n}\dot{\Psi}_{n}\rangle)\\ 
    &+ \Psi_{n}^2\dot{\Psi}_{n}-\langle\Psi_{n}^2\dot{\Psi}_{n}\rangle)
    = g_n \cos \Omega t
    \end{split}
    \label{eq:fast_dynamics}
\end{equation}

According to the inertial approximation \cite{blekhman2000vibrational}, the second derivative of the fast variable dominates its dynamics ($\ddot{\Psi}_{n} \gg \dot{\Psi}_{n}, \Psi_{n}, \dots$). Equation \eqref{eq:fast_dynamics} simplifies to:
$$
\ddot{\Psi}_{n} \approx g_{n} \cos \Omega t
$$
Integrating twice yields the solution for $\Psi_n$:
\begin{equation}
\Psi_{n} = -\frac{g_{n}}{\Omega^{2}}\cos\Omega t=f_n\cos\Omega t,
\end{equation}
where $f_n=\dfrac{g_n}{\Omega^2}$ represents the high frequency signal strength, which is assumed to be of $\mathcal{O}(1)$. From this, the time averages of different moments can be calculated as:

\begin{equation}
\langle\Psi_{n}^{2}\rangle = \frac{g_{n}^{2}}{2\Omega^{4}}\quad\mathrm{and}\quad\langle\Psi_{n}^{3}\rangle = 0
\end{equation}

Substituting these moments back into the averaged Eq.\eqref{eq:effective_average_slow}, assuming $\langle g_n \cos \Omega t \rangle = 0$, the effective equation for the slow dynamics can be recast as:

\begin{equation}
\begin{split}
   \ddot{U}_{n}&+\epsilon\gamma\dot{U}_{n}+(\omega_{0}^{2}+\epsilon h \cos \omega_{p}t)U_{n} + \frac{1}{2}\epsilon d(U_{n+1}-2U_{n}+U_{n-1})\\
   &+ \epsilon\alpha\left(\dfrac{3}{2}f_n^2U_n+ U_{n}^{3}\right)+\epsilon\eta\left(U_n^2+\dfrac{f_n^2}{2}\right)\dot{U}_{n}=0 
   \end{split}
   \label{eq:effective_slow_1}
\end{equation}

Combining the terms with $\dot{U}_n$, we obtain the final form:
\begin{equation}
\begin{split}
   \ddot{U}_{n}&+\gamma_{eff}\dot{U}_{n}+(\tilde{\omega}^{2}+\epsilon h \cos \omega_{p}t)U_{n} + \frac{1}{2}\epsilon d(U_{n+1}-2U_{n}+U_{n-1})\\
   &+\epsilon\alpha U_{n}^{3}+\epsilon\eta U_n^2\dot{U}_n=0 
   \end{split}
   \label{eq:effective_slow_final}
\end{equation}
where
\begin{equation}
  \gamma_{eff,n}=\left(\gamma+\dfrac{\eta f_n^{2}}{2}\right)\quad \mathrm{and} \quad
\tilde{\omega_{n}}^{2}=\omega_0^2+\dfrac{3\epsilon\alpha f_n^2}{2} 
\label{eq:effective_gama_omega}
\end{equation}

Thus, the primary effect of the uniform HF additive forcing is to \emph{shift the natural frequency} and to add a small correction to the \emph{linear damping} through the HF drive strength $f_n^2$. The cubic coefficient \(\alpha\) remains (to leading order) unchanged, but the effective linear term is modified by coupling between \(U\) and the averaged fast variance \(\langle\Psi^2\rangle\).

\subsection{Deriving the flow equations}

The dispersion relation, the modes of oscillation, and their tuning through the HF strength $f_n$ can be analytically obtained by means of applying multi-time scale perturbation analysis to Eq.~\eqref{eq:effective_slow_final}. For this purpose let us introduce a dimensionless time scale $\omega_p t = \tau$. So that equation Eq.~\eqref{eq:effective_slow_final} becomes:

\begin{equation}
\begin{split}
U_n'' &+ \epsilon \Gamma_1 U_n' + 
\left( \frac{\tilde{\omega_{n}}^2}{\omega_p^2} + \epsilon H \cos \tau \right) U_n \\
+& \frac{1}{2}\epsilon D (U_{n+1} - 2U_n + U_{n-1})
+\dfrac{1}{2} \epsilon \Gamma_2 f_n^2 U_n^2 U_n' + \epsilon \beta U_n^3 = 0
\end{split}
\label{eq:effective_slow_perturbed}
\end{equation}

where the prime ($'$) denotes differentiation with respect to $\tau$ and:
\[
\Gamma_1 = \frac{\gamma}{\omega_p}, \quad 
D = \frac{d}{\omega_p^2}, \quad
\Gamma_2 = \frac{\eta}{\omega_p}, \quad 
\beta = \frac{\alpha}{\omega_p^2}, \quad H=\dfrac{h}{\omega_p^2}
\]

Further, the leading contribution of parametric resonance occurs when the parametric frequency is twice the natural frequency, i.e.,
\begin{equation}
\omega_p = 2\tilde{\omega_{n}} + \epsilon \tilde{\sigma}_p, \quad
\frac{\tilde{\omega_{n}}^2}{\omega_p^2} = \dfrac{1}{4} - \epsilon \sigma_p
\label{eq:detuning_parameter}
\end{equation}

where $\sigma_p=\dfrac{\tilde{\omega_{n}}\tilde{\sigma}_p}{\omega_p^2}$ is the detuning parameter. Now Eq.~\eqref{eq:effective_slow_perturbed} becomes:

\begin{equation}
\begin{split}
U_n'' &+ \epsilon \Gamma_1 U_n' + 
\left( \frac{1}{4} + \epsilon H \cos \tau \right) U_n\\
&+ \frac{1}{2} D (U_{n+1} - 2U_n + U_{n-1}) 
+\dfrac{1}{2} \epsilon \Gamma_2 f_n^2 U_n^2 U_n' + \epsilon \beta U_n^3-\epsilon \sigma_p = 0 
\end{split}
\label{eq:effective_slow_detuned}
\end{equation}

\begin{figure}[ht!]
     \subcaptionbox{Variation of mode with resonant drive strength $g_{res}$}{\includegraphics[scale=0.29]{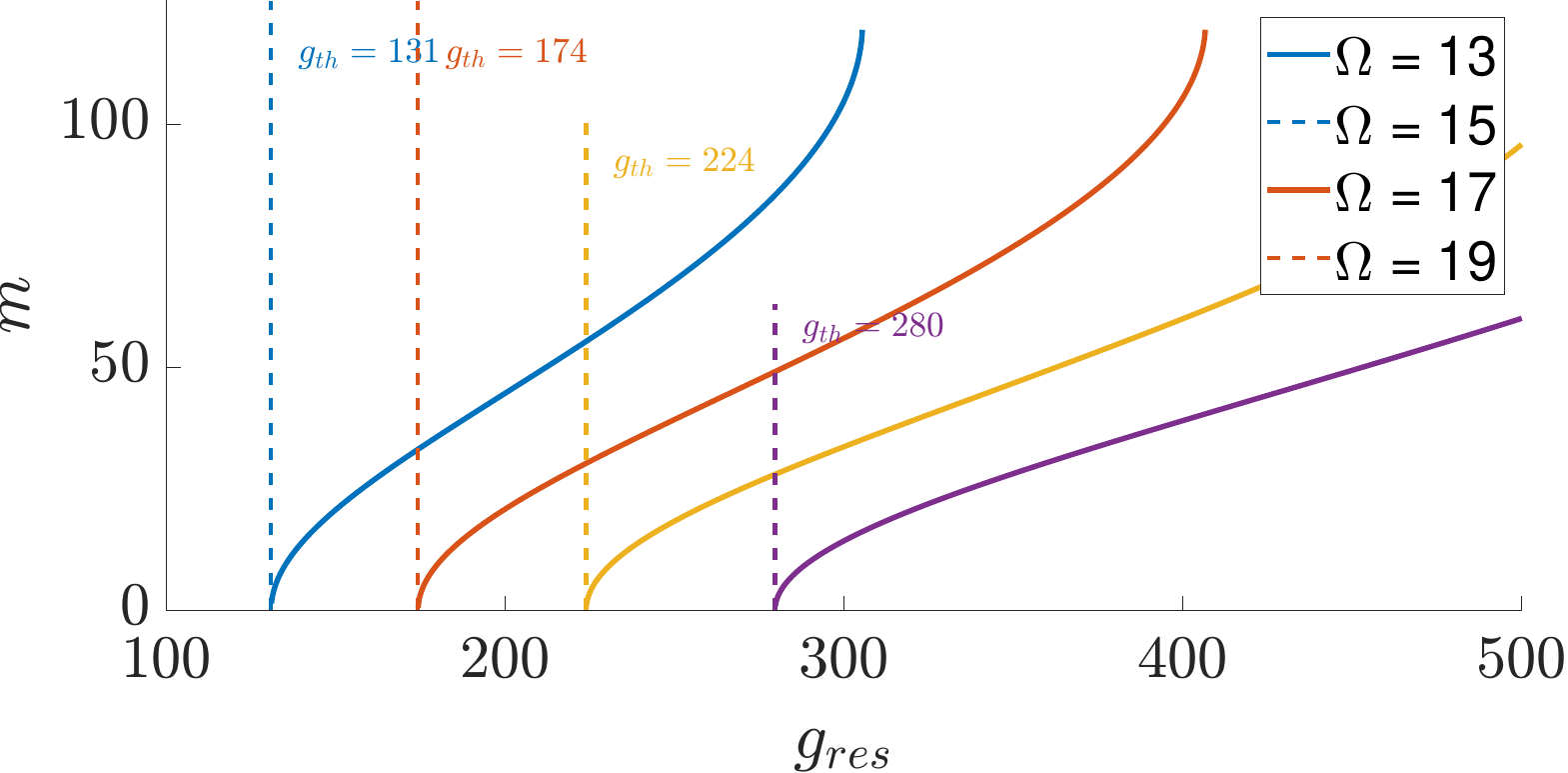}}
    \subcaptionbox{Variation of mode with $\Omega$}{\includegraphics[scale=0.38]{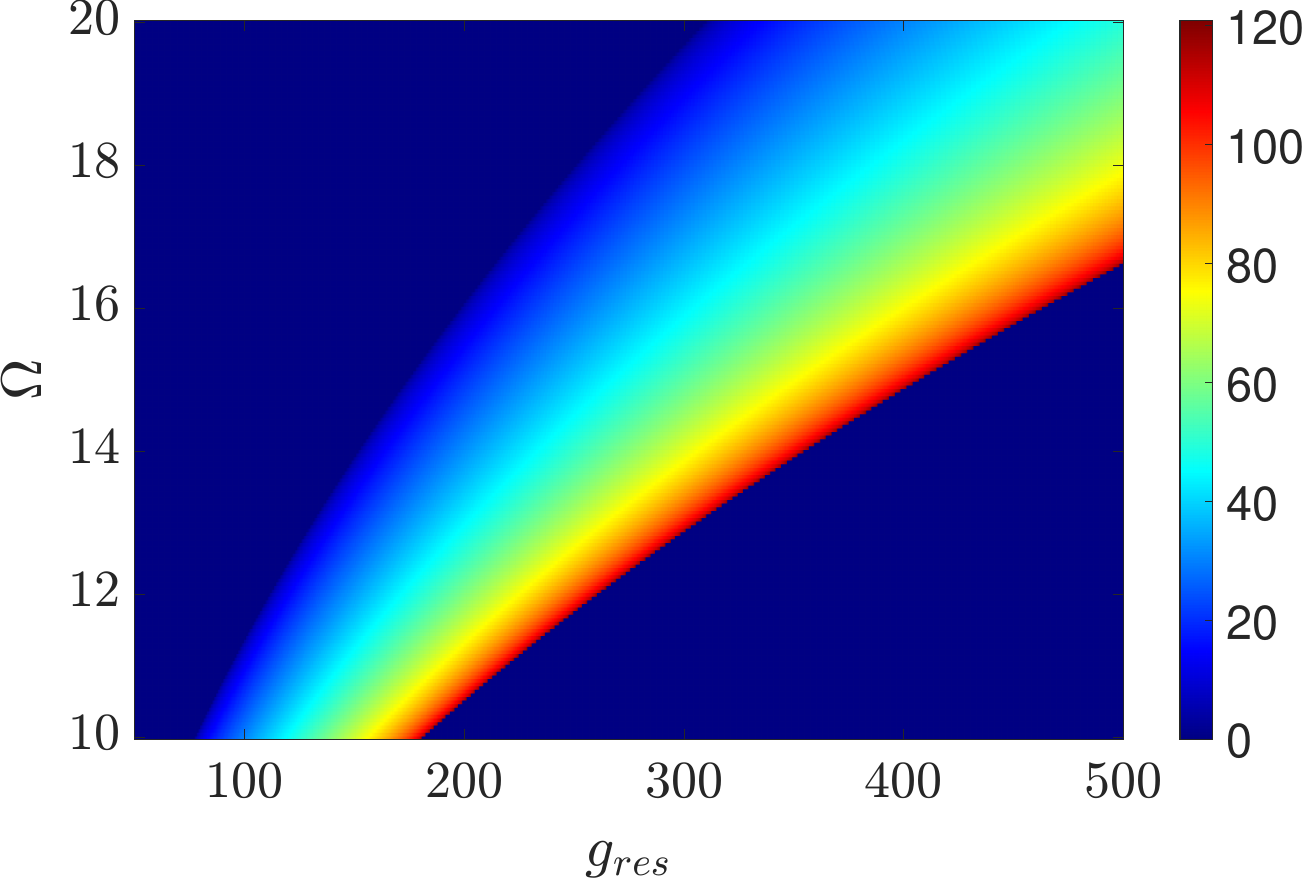}}
    \subcaptionbox{Variation of mode with N}{\includegraphics[scale=0.38]{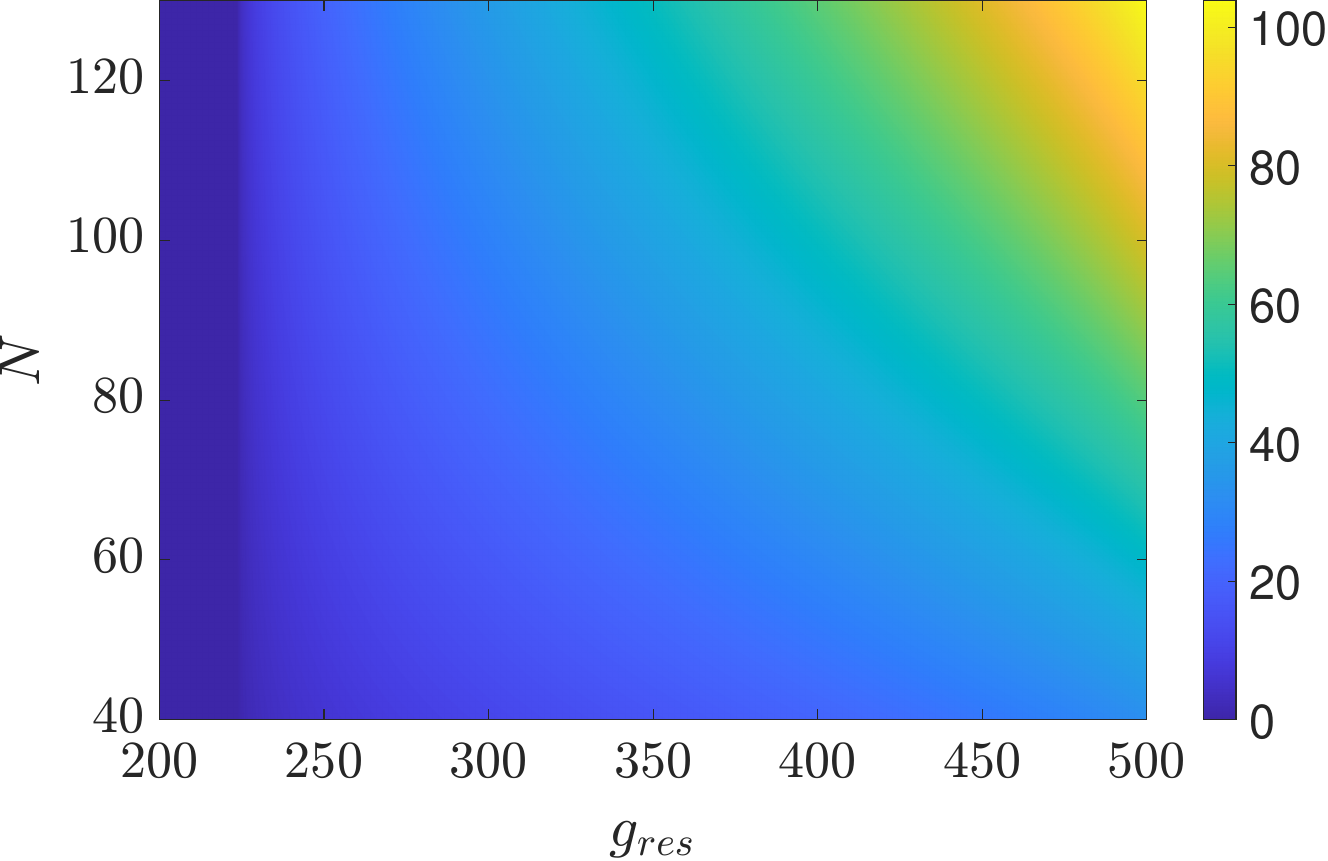}}

    \caption{FIG. 2. HF–driven tuning of collective mode selection in the nonlinear resonator array. 
(a) Variation of the resonant mode index \(m\) with the HF resonant drive amplitude \(g_{res}\) by Eq.~\eqref{eq:g_peak}. Also describing the 
minimum HF threshold $g_{th}$ required to activate bulk modes for different drive frequencies \(\Omega\). 
(b) Mode-index colormap in the \(g_{res}\text{--}\Omega\) plane by Eq.~\eqref{eq:mode_index}, showing how the HF-induced dispersion shift selectively brings the specific modes into the parametric resonance window. 
(c) Dependence of the resonant mode number on the HF resonant amplitude for arrays of different sizes \(N\), 
showing that the percentile mode position \(m/N\) remains nearly invariant across system sizes, 
highlighting the collective nature of mode selection.The parameters are taken as
$\epsilon = 0.1;\omega_0= 0.4;\omega_p= 1.0;d= 2.0;\alpha= 1.0;\gamma=2.0;\eta=2.0;h=3.9$.}

    \label{fig:mode_index_analytical}
\end{figure}

Since the nonlinearity and the parametric excitation are weak (scaled by $\epsilon$), the asymptotic expansion of $U_n$ is formulated using the method of multiple time scales. Defining the slow time $\tau_1=\epsilon\tau$ and the fast time $\tau_0=\tau$, the expansion reads:

\begin{equation}
U_n(t) = U_n^{(0)} + \epsilon U_n^{(1)} + \ldots~~~~\text{and}~~~ \dfrac{d}{d\tau}=\dfrac{\partial}{\partial\tau_0}+\epsilon\dfrac{\partial}{\partial\tau_1}+\ldots
\label{eq:perturbative_expansion}
\end{equation}

Substituting Eq.~\eqref{eq:perturbative_expansion} into Eq.~\eqref{eq:effective_slow_perturbed} and equating the coefficients of like powers of $\epsilon$ gives:  At $\mathcal{O}(\epsilon^0)$:
\begin{equation}
  U_n^{(0)''} + \frac{1}{4} U_n^{(0)} = 0
  \label{eq:zeroth_order_eq}
\end{equation}

The zeroth-order solution is described by a superposition of standing wave modes with slowly varying amplitudes:
\begin{equation}
    U_n^{(0)} = \frac{1}{2} \sum_m A_m(\tau_1) \sin(n q_m) e^{i \tau_0 / 2} + c.c.
    \label{eq:zeroth_order_sol}
\end{equation}
Here $q_m=\dfrac{m\pi}{N+1}$ is the dimensionless wavenumber with $m$-th linear normal mode ($m=1,2...N$).The expression of the wavenumber arises from the fixed boundary conditions imposed at the ends of the array ($U_0=U_{N+1}=0$).

Now at $\mathcal{O}(\epsilon^1)$ Eq.~\eqref{eq:effective_slow_detuned} yields

\begin{equation}
\begin{split}
     U_n^{(1)''} + \frac{1}{4} U_n^{(1)}=&-2\dfrac{\partial^2 U_n^{(0)}}{\partial\tau_1\partial\tau_0}-\Gamma_1 U_n^{(0)'}-H \cos \tau_0 \, U_n^{(0)}\\
     &-\frac{D}{2} (U_{n+1}^{(0)} - 2U_n^{(0)} + U_{n-1}^{(0)})\\
     &-\beta U_n^{(0)^3}-\dfrac{1}{2}\Gamma_2 f_n^2 U_n^{(0)^2} U_n^{(0)'}+\sigma_{p}U_n^{(0)}
     \end{split}
     \label{eq:first_order_eq}
\end{equation}

Substituting Eq.~\eqref{eq:zeroth_order_sol} into Eq.~\eqref{eq:first_order_eq} and evaluating the terms separately (Appendix~\ref{app:A}), we arrive at the complex amplitude flow equation by equating all secular terms proportional to $\exp(i\tau_0/2)$ to zero:

\begin{equation}
\begin{split}
-i \frac{\partial A_m}{\partial \tau_1}
-& \frac{i \Gamma_1}{2} A_m
+ 2D \sin^2 \left( \frac{q_m}{2} \right) A_m
- \frac{H}{2} A_m^*\\
-& \left( \frac{3 \beta}{16} + \frac{i \Gamma_2f_n^2}{64} \right)
\sum_{j,k,l} A_j A_k A_l^* \Delta_{j,k,l,m}
+ \sigma_p A_m = 0
\label{eq:amp_flow}
\end{split}
\end{equation}

Where we use the $\Delta$ function as defined in \cite{lifshitz2003response} as 

\begin{equation}
\begin{split}
\Delta^{(1)}_{jkl;m} &= 
\delta_{-j+k+l,m}
- \delta_{-j+k+l,-m}
- \delta_{-j+k+l,2(N+1)-m} \\
&\quad + \delta_{j-k+l,m}
- \delta_{j-k+l,-m}
- \delta_{j-k+l,2(N+1)-m} \\
&\quad + \delta_{j+k-l,m}
- \delta_{j+k-l,-m}
- \delta_{j+k-l,2(N+1)-m} \\
&\quad - \delta_{j+k+l,m}
+ \delta_{j+k+l,2(N+1)-m}
- \delta_{j+k+l,2(N+1)+m}.
\end{split}
\label{eq:delta_fn}
\end{equation}

The function $\Delta_{jkl;m}^{(1)}$ acts as a \emph{selection rule} that governs the energy exchange through the nonlinear damping and the cubic stiffness. The first Kronecker delta in each line represents the direct momentum conservation, and the subsequent two is responsible for the so-called \emph{Umklapp condition} (where only lattice momentum is conserved) due to the discrete fixed boundaries.
It is well known that for a single oscillator, the steady state amplitude response to the parametric response can be written as $A_m=a_m e^{i\sigma_p\tau_1}$, which can be substituted into Eq.~\eqref{eq:zeroth_order_sol} to obtain

\begin{equation}
    U_n^{(0)} = \frac{1}{2} \sum_m a_m (\tau_1) \sin(n q_m)e^{i\left(\dfrac{1}{2}+\epsilon\sigma_p\right)\tau_0} + c.c.
    \label{eq:zeroth_order_sol_2}
\end{equation}
where all the modes are oscillating at half of parametric frequency that is determined by Eq.~\eqref{eq:detuning_parameter}.Then the steady state solution is obtained by putting $A_m$ into Eq.~\eqref{eq:amp_flow}, yields the time independent amplitude equation in terms of the original system parameters,i.e., $\left(\dfrac{\partial a_m}{\partial \tau_1}=0\right)$ as:

\begin{equation}
\begin{split}
\Bigg( \sigma_p +& 2\frac{d}{\omega_p^2} \sin^2 \!\Big( \frac{q_m}{2} \Big)
- \frac{i \gamma}{2\omega_p} \Bigg) a_m
- \frac{h}{2\omega_p^2} a_m^*\\
-& \left( \frac{3 \alpha}{16\omega_p^2} + \frac{i \eta f_n^2}{64\omega_p} \right)
\sum_{j,k,l} a_j a_k a_l^* \Delta_{j,k,l,m} = 0
\end{split}
\label{eq:amp_flow_time_independent}
\end{equation}

\begin{figure}[ht!]
    \centering
    \includegraphics[scale=0.38]{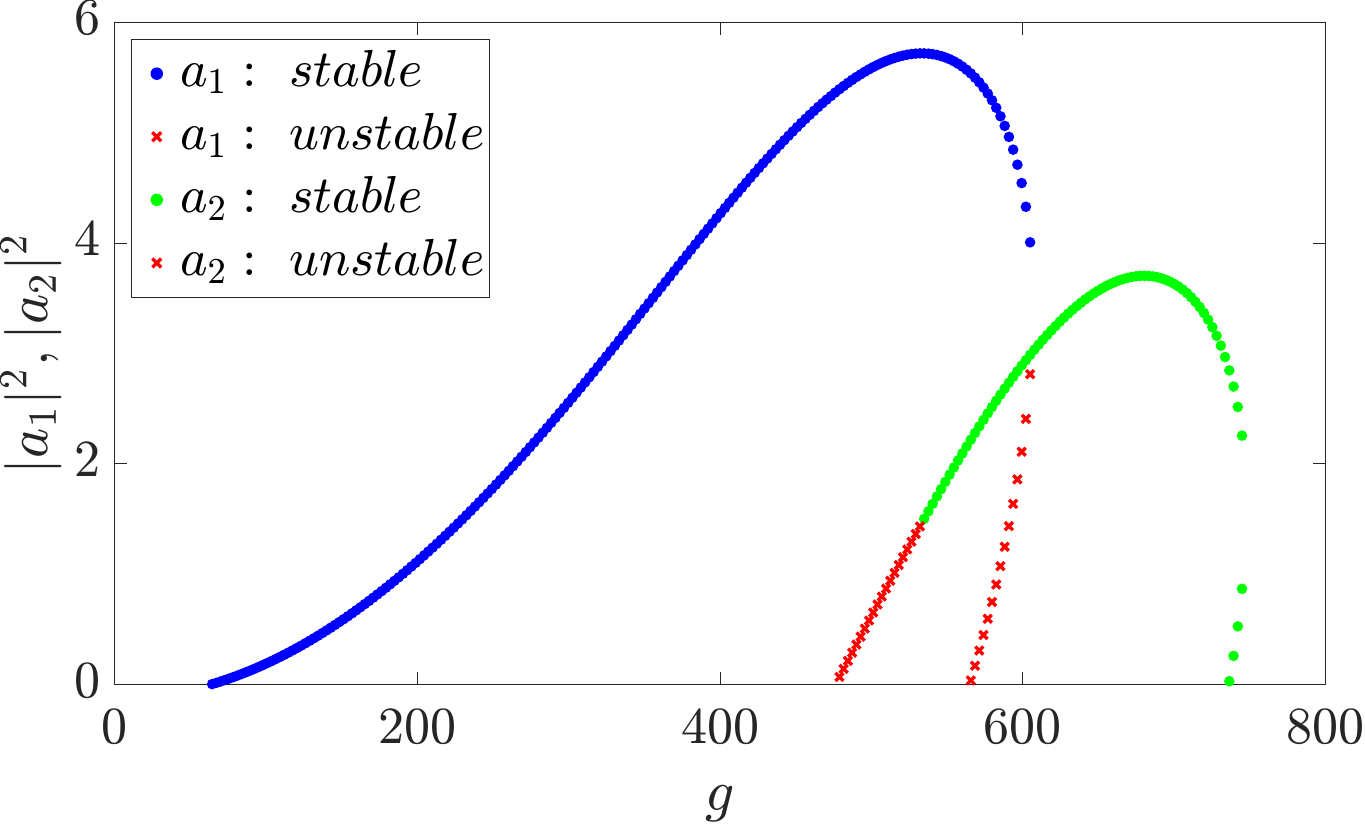}
    \caption{Response intensity $a_1$ and $a_2$ as a function of HF drive strength $g$ for two oscillators $N=2$ is shown.The stable and unstable single-mode solutions are obtained from Eqs.~\eqref{eq:a_1_single_mode} and \eqref{eq:a_2_single_mode}. The parameters are fixed at $\epsilon = 0.1;\omega_0= 0.4;\omega_p= 1.0;d= 2.0;\alpha= 1.0;\gamma=2.0;\eta=2.0;h=6.0$.}
    \label{fig:two_oscillator_single_mode}
\end{figure}

 This is the main perturbative result which replaces the $N$ coupled differential equations Eq.~\eqref{eq:effective_slow_perturbed} for the effective displacement $U_n$ to $N$ coupled algebraic equations for \emph{time-independent} modes amplitudes $a_m$. The presence of the fast forcing term $f_n$ explicitly coupled with $\eta$ and implicitly via $\tilde{\omega_n}$ in $\sigma_p$ implicates the fact that the mode amplitudes are now functionally dependent on the HF signal strength $f_n$ or the fast forcing drive $g$. The first two terms in Eq.~\eqref{eq:amp_flow_time_independent} describe that the linear resonance response is obtained for

 \begin{equation}
\delta_p =\sigma_p + 2\frac{d}{\omega_p^2} \sin^2 \!\left( \frac{q_m}{2} \right)=0
\label{eq:effective_detuning}
 \end{equation}
 rather only $\sigma_p=0$.The term $\delta_p$ describes the effective frequency mismatch for the particular mode, emerging from the combined effect of onsite detuning $\sigma_p$ and the coupling constant $d$. Then, by Eq.~\eqref{eq:detuning_parameter}, replacing back $\sigma_p$, the mode dispersion relation for the parametric response (at $\omega_p/2$) is obtained as:

\begin{equation}
    \tilde{\omega}_n^2=\dfrac{\omega_p^2}{4}+2\epsilon d\sin^2(\dfrac{q_m}{2})
    \label{eq:mode_dispersion_relation}
\end{equation}

The above equation gives the fundamental connection between the effective frequency and the modes of oscillation. This gives the freedom to express the above equation in terms of HF strength as

\begin{equation}
    g_{\text{res}} =\sqrt{\frac{2\Omega^4}{3\epsilon\alpha} \left[ \frac{\omega_p^2}{4} - \left( \omega_0^2 - 2\epsilon d\sin^2\left(\frac{q_m}{2}\right) \right) \right]}
    \label{eq:g_peak}
\end{equation}

This relation gives the quantitative description of how the modes resonating with a specific frequency can be externally tuned by an HF drive. The term $g_{\text{res}}$ denotes the peak of the collective response, that indicates the maximum energy pumping through the parametric drive is transferred through specific mode $q_m$. The mode index can reversibly expressed as:

 \begin{equation}
     m(g_{\text{res}}) = \frac{2(N+1)}{\pi} \arcsin \left[ \sqrt{ \frac{1}{2\epsilon d} \left( \frac{3\epsilon\alpha g_{\text{res}}^2}{2\Omega^4} - \left[ \frac{\omega_p^2}{4} - \omega_0^2 \right] \right) } \right]
     \label{eq:mode_index}
 \end{equation}

\begin{figure*}[ht!]
  \centering
  \includegraphics[width=\textwidth]{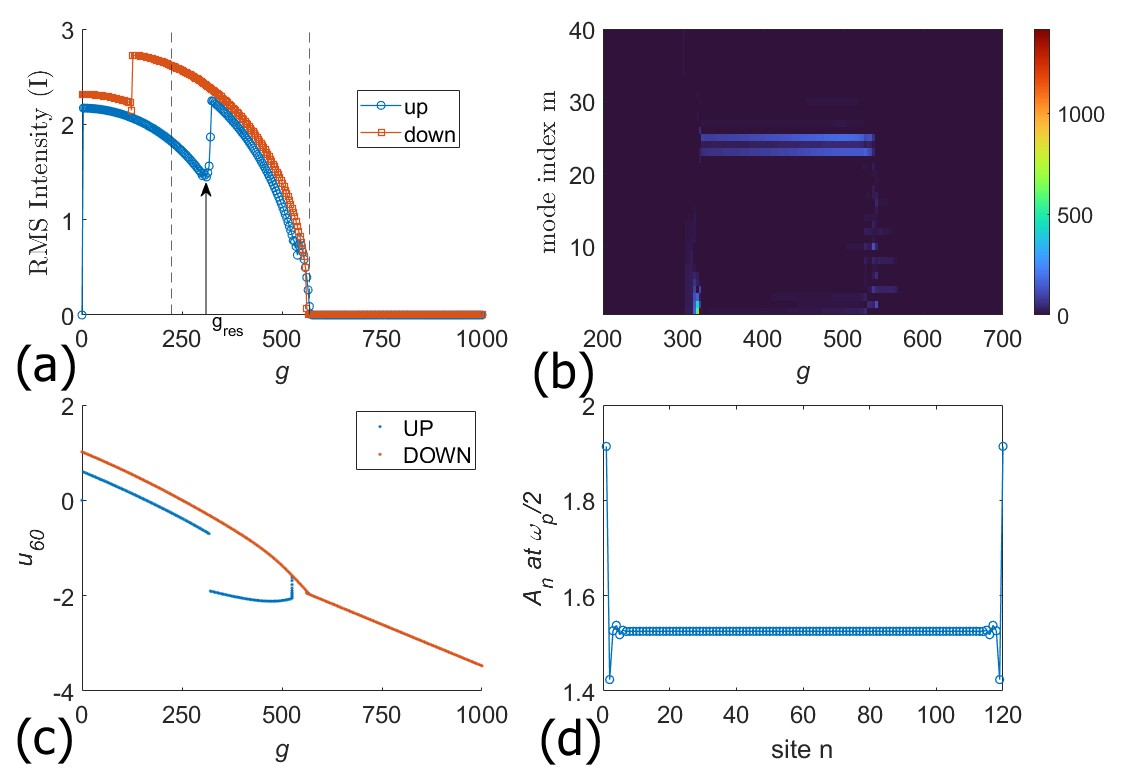}
  \caption{The collective parametric response in the nonlinear resonator array ($N=120$) as a function of the high-frequency drive amplitude ($g$) is validated numerically. (a) RMS intensity (I) against $g$. The critical hysteretic region is depicted in the plot. While  the blue line (up sweep) jumps up at the threshold ($g_{up}=g_{res} \approx 290$),the down sweep occurs at nearly $g=130$.(b) Shows the energy intensity localized in each normal mode ($m$) throughout the $g$ sweep. At near resonant drive $g_{res}=290$, the response is highly selective, confining almost all energy to a narrow band of bulk propagating modes ($m \approx 20-25$), confirming the close approximation of mode tuning mechanism through Eq.\eqref{eq:mode_index}. (c) The bifurcation plot of the mean displacement of the central node ($n=60$) is non-zero, suggesting the existence of jump stability during up sweep near $g_{res}=290$.The dynamically oscillating state is represented by the area between the UP and DOWN curves.  (d) The amplitude distribution at the maximum response peak $g_{res}=290$ across all lattice sites ($n$).  The near-uniform amplitude validates the collective nature of the resonance by confirming the selective excitation of a \textit{bulk propagating mode} ($m=31$) with minimal boundary effects. The other related parameters are used as:$\epsilon = 0.1;\omega_0= 0.4;\omega_p= 1.0;d= 2.0;\alpha= 1.0;\gamma=2.0;\eta=2.0;h=3.9$}
  \label{fig:collective_resonance}
\end{figure*}

\begin{figure*}[ht!]
  \centering
  \includegraphics[width=\textwidth]{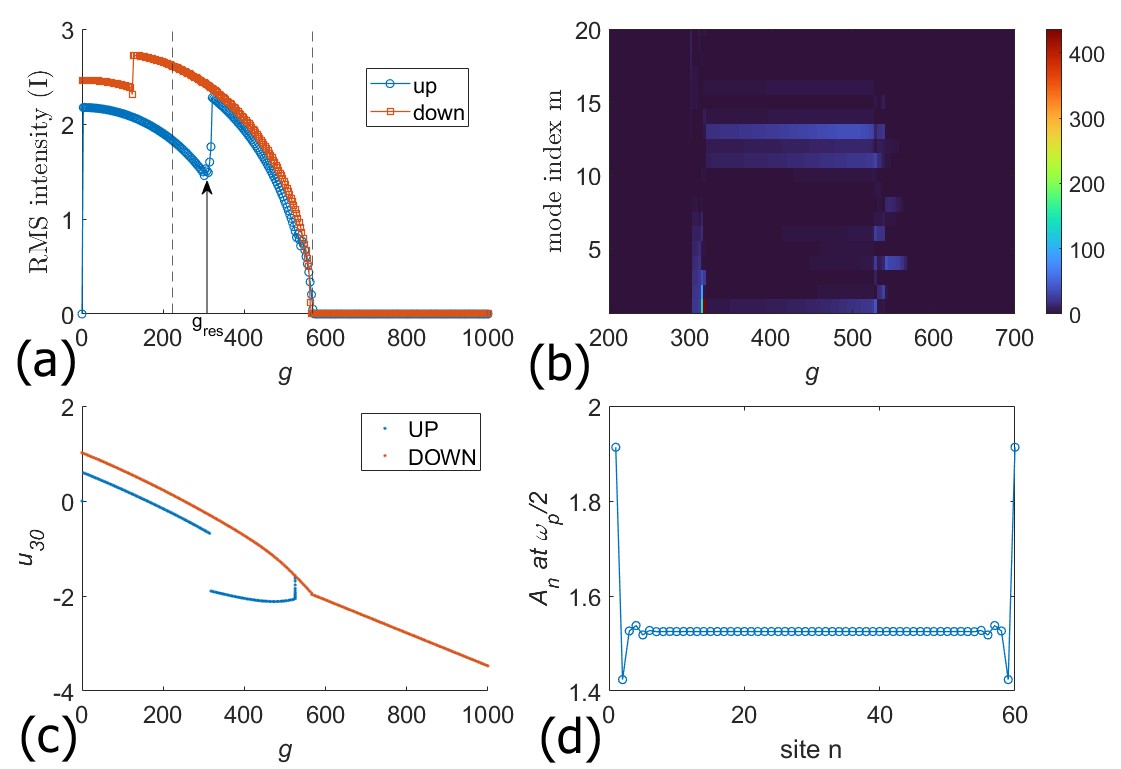}
  \caption{The collective parametric response in the nonlinear resonator array ($N=60$) as a function of the high-frequency drive amplitude ($g$) is validated numerically. (a) RMS intensity (I) against $g$. The critical hysteretic region is depicted in the plot. While the blue line (up sweep) jumps up at the threshold ($g_{up}=g_{res} \approx 290$),the down sweep occurs at nearly $g=130$.(b) Shows the energy intensity localized in each normal mode ($m$) throughout the $g$ sweep. At near resonant drive $g_{res}=290$, the response is highly selective, confining almost all energy to a narrow band of bulk propagating modes ($m \approx 10-15$), confirming the close approximation of mode tuning mechanism through Eq.\eqref{eq:mode_index}. (c) The bifurcation plot of the mean displacement of the central node ($n=30$) is non-zero, suggesting the existence of jump stability during the up sweep near $g_{res}=290$. The dynamically oscillating state is represented by the area between the UP and DOWN curves.  (d) The amplitude distribution at the maximum response peak $g_{res}=290$ across all lattice sites ($n$).  The near-uniform amplitude validates the collective nature of the resonance by confirming the selective excitation of a \textit{bulk propagating mode} ($m=16$) with minimal boundary effects. The other related parameters are used as:$\epsilon = 0.1;\omega_0= 0.4;\omega_p= 1.0;d= 2.0;\alpha= 1.0;\gamma=2.0;\eta=2.0;h=3.9$}
  \label{fig:collective_resonance2}
\end{figure*}

Eq.~\eqref{eq:mode_index} determines the specific spatial mode excited at the resonance peak, a relationship that can be directly validated through numerical investigation. By solving the full original dynamics Eq.~\eqref{eq:EOM} to determine the critical drive amplitude $g_{\text{res}}$, the corresponding mode index $m$ can be explicitly recovered. This index characterizes the collective behavior of the array: a value of $m=1$ indicates an ideal, coherent collective oscillation (acoustic limit), while $m \to N$ corresponds to short-wavelength, anti-phase dynamics (optical limit) often associated with the onset of localization. The variation of mode index $m$ with $g_{res}$ is plotted in Fig.\ref{fig:mode_index_analytical}(a) for different values of high frequency drive $\Omega$ by Eq.~\eqref{eq:mode_index}, which reveals that for a particular value of $\Omega$, a minimum threshold $g_{th}$ HF drive is needed to excite the modes, for example for $\Omega=17$ the threshold value is $g_{th}=174$. In Fig.\ref{fig:mode_index_analytical}(b), the mode colormap is synthesized in the $g-\Omega$ plane.This plot defines the operational phase space of the array, which can be useful to design, say, a narrow-band filter targeting $m=80$; a controller needs only to locate the necessary $g_{\text{res}}$ value on the $m=80$ horizontal line for the chosen $\Omega$. Finally, in Fig.\ref{fig:mode_index_analytical}(c), the number of mode participating in resonant vibration is mapped in ($N-g_{res}$) plane. To achieve highly localized mode oscillation, the HF drive should be increased proportionally, while for the collective response, the percentile mode selection ($m/N$) is nearly same for a particular resonant drive across different array populations.   \\
In Fig.\ref{fig:two_oscillator_single_mode}, the single-mode solution for two oscillators is described. The two elliptical single-mode solution branches, for $N=2$, are derived from the Eq.~\eqref{eq:amp_flow_time_independent}(See Appendix: \ref{app:B}, Eqs.~\eqref{eq:a_1_single_mode} and \eqref{eq:a_2_single_mode}).All the other parameters are fixed as in the previous cases,except the parametric strength is raised to the value $h=6.0$ instead of $3.9$. For the lower value of $h$ unstable branch of $a_2$ does not appear distinctly. The primary motivation of the article is to demonstrate the active tuning method by an HF drive $g$, bypassing the conventional method of tuning the pump frequency $\omega_p$, which can drive the system to resonate with a parametric drive and also activate the selective mode. Given the focus on that, the complex multi-mode behavior of the system is not presented in Fig.\ref{fig:two_oscillator_single_mode} for brevity. Such dual mode competition would necessitate the higher-order stability analysis and is reserved for future research direction.

\section{RESPONSE AND MODE SELECTION: NUMERICAL RESULTS}

The numerical results presented in this section focus on two distinct methods. The first is to extract the effective slow response of the array [Eq.~\eqref{eq:EOM}] subjected to the external HF drive, following the pioneering work of McLintock and Landa \cite{landa2000vibrational}. The second part describes how this effective motion collectively respond to a parametric excitation by focusing its energy to a particular normal mode of the system when tuned by the HF drive strength. This analysis is crucial for understanding the dynamics of lattice response and micro or nano resonators, as established by Cross et al. \cite{lifshitz2003response}. At first, the response of the system is taken at the frequency $\omega_p/2$ (for leading parametric resonance) as the rms average by evaluating the Fourier sine and cosine components as
\begin{eqnarray}
    Q_C=\dfrac{1}{T}\int_0^T u_n\cos(\dfrac{\omega_p t}{2}) \diff t\\
    Q_S=\dfrac{1}{T}\int_0^T u_n\sin(\dfrac{\omega_p t}{2}) \diff t
\end{eqnarray}
Then by defining the quality factor $Q_n(g)=\dfrac{\sqrt{Q_C^2+Q_S^2}}{h}$, which determines the enhancement of the array response to the input weak parametric pumping as the HF signal strength $g$ is varied. Finally the the intensity of the whole array as a collective response is defined by 
\begin{equation}
    I(g)=\dfrac{1}{N}\sum_{n=1}^N\dfrac{Q_n}{\sqrt{2}}
    \label{eq:numerical_intensity}
\end{equation}

In Fig.\ref{fig:collective_resonance}(a), the response of the array is plotted against the drive strength $g$. The two-way sweep (up and down) of $g$ has been employed to capture the hysteresis effect. From the figure, the peak value $g_{\text{res}}\approx 290$ is where the UP transition occurs. At this value of the HF drive, the collective response of the system to the parametric drive $h\cos(\omega_p t)$ is maximum, and the energy is localized to a particular mode of oscillation $q_m$, which is determined by the expression Eq.~\eqref{eq:mode_index}. For $N=120$ oscillators, and the resonant drive $g_{\text{res}}=290$, the mode index is approximately found out to be $m=31$ from Eq.~\eqref{eq:mode_index}. Evidently, this low mode index refers to \emph{bulk propagation mode}, indicating the collective response of the array. The resonant energy transfer to $q_{31}$ is also validated by the numerical in Fig.\ref{fig:collective_resonance}(b), which clearly depicts the energy band around $m=25$. The concentration of energy in $m=20$ to $m=25$ confirms that the frequency tuning is highly selective. In the simulations, the response therefore appears as a narrow mode packet rather than a single discrete mode index, which is typical for a finite array. The slight discrepancy in the value of the selective resonant mode obtained from the analytical result Eq.~\eqref{eq:mode_index} and from the numerical is expected. While the analytical results are derived after the application of a successive perturbation scheme (first averaging, then multi time scale--with the approximation truncated at first order), the numerical contain all the time scales. Also, the inclusion of perturbation only up to 1st order contributes to the mismatch, and higher order corrections would further reduce this mismatch and yield analytically more accurate mode estimates. For the specific combination of $\omega_0$ and $\omega_p$ used, the HF drive $g$ successfully tunes the total array stiffness ($\omega_{\mathrm{eff}}$) to bring the near $\mathbf{25^{th}(m/N)\text{-percentile mode}}$ approximately into the parametric resonance window. This visualization confirms the collective nature of the tuning, demonstrating that the energy is channeled into coordinated bulk waves (modes $m=20$ to $25$) that span the entire array, rather than simple single-site oscillations. In Fig\ref{fig:collective_resonance}(c), the Poincar\'e map of steady-state central mode ($u_{60}$) is shown, which describe the change in stability branch at nearly $g=290$. The uniformity of the mode-amplitude over different site at the resonant oscillation is depicted in Fig\ref{fig:collective_resonance}(d), confirming that the selected mode ($m=25$) successfully excites a bulk propagating wave where the energy is distributed uniformly.\\ 
The simulation is performed over another set of $N=60$ oscillator to check the robustness of this tuning capability through the HF drive $g$, which is displayed in Fig.\ref{fig:collective_resonance2}. The result confirms nearly unchanged the hysteresis window and the percentile mode selection ratio $m/N$ which is about $m=12$. Fig.\ref{fig:collective_resonance2}(b) shows the concentrated energy band near $m=12$ to $m=15$ for the same specific value of $g_{res}=290$, while the analytical result for $N=60$ (Eq.~\eqref{eq:mode_index}) predicts the value is about $m=16$, which is in close agreement. The bulk mode wave propagation can also be visualized in  Fig.\ref{fig:collective_resonance2}(d), where the resonant oscillation is uniform across the sites. The analytical prediction of the percentile mode index \(m/N\) for \(N=120\) and \(N=60\) is approximately \(25\%\), corresponding to the analytically obtained modes \(m=31\) and \(m=16\). The numerically extracted mode indices, \(m \approx 25\) and \(m \approx 12\), yield a percentile value of about \(20\%\), which remains in close agreement with the analytical prescription for selective mode activation.

\section{Conclusions}

This work develops a dynamic mechanism to control the collective response of a self-sustained coupled parametric oscillator array by an HF drive. By assuming a standing wave mode solution and using the multi-time scale perturbation method, the effective-averaged dynamics of the array are demonstrated, where the onsite stiffness and damping are renormalized due to HF drive. A controllable shift in the dispersion relation induced by the HF signal thereby enables the selective bulk mode wave propagation, demonstrating the mode-specific controlled response to the weak parametric excitation. The analytical predictions are validated through the numerical simulations, revealing sharp mode selection and coherent \emph{collective vibrational resonance} across the array. The general steady state dynamics of the modes and their connection to HF parameters are also presented with a detailed derivation of the two-mode solution. Finally, this work can be further expanded to find the effect of HF drive in a continuous regime and structural change in the characteristics of intrinsic localized modes (ILM) as a future research endeavor.

\begin{acknowledgments}
% Acknowledgments (edit as appropriate)
SR is grateful to the Director, Prof.Satyajit Chakrabarti, for providing the facility to pursue this research work at the Institute of Engineering \& Management (IEM Kolkata), and to Prof. K.P. Ghatak for his encouragement and guidance. SR is also thankful to Dr.Dhrubajyoti Biswas (National Brain Research Centre, Gurgaon) for the discussions on the numerical simulations.
\end{acknowledgments}

\bibliographystyle{unsrt}
\bibliography{refeences_vr_array}

\appendix

\section{\label{app:A} Derivation of complex amplitude flow equation}
The RHS of Eq~\eqref{eq:first_order_eq} has several terms, which are analyzed as:
The derivative term:
\begin{equation}
    \frac{\partial^2 U_n^{(0)}}{\partial \tau_1 \partial \tau_0}
= \frac{i}{4} \sum_m \dfrac{\partial A_m}{\partial\tau_1} \sin(n q_m) e^{i \tau_0 / 2} + c.c.
\label{eq:second_derivative}
\end{equation}

The parametric term gives:
\begin{equation}
    H \cos \tau_0 \, U_n^{(0)} = 
\frac{H}{4} \sum_m A_m \sin(n q_m) e^{i \tau_0 / 2} + c.c.
\label{eq:parametric_term}
\end{equation}

The coupling term yields:

\begin{equation}
\begin{split}
\dfrac{D}{2} (U_{n+1}^{(0)} -& 2U_n^{(0)} + U_{n-1}^{(0)}) \\
=& -D \sum_m \sin^2 \!\left( \frac{q_m}{2} \right)
\sin(n q_m) (A_m e^{i \tau_0 / 2} + c.c.) 
\end{split}
\label{eq:coupling_term}
\end{equation}

The cubic nonlinearity reads:

\begin{equation}
\begin{split}
   \beta U_n^{(0)^3} &= 
\frac{\beta}{8} \sum_{j,k,l} 
\sin(n q_j) \sin(n q_k) \sin(n q_l)
\big\{(A_j e^{i \tau_0 / 2} + c.c.)\\
&(A_k e^{i \tau_0 / 2} + c.c.)
(A_l e^{i \tau_0 / 2} + c.c.)\big\}\\
&=\frac{\beta}{32}\sum_{j,k,l} 
\big\{ 
\sin[n(-q_j + q_k + q_l)]+\sin[n(q_j + q_k - q_l)] \\
&+
\sin[n(q_j - q_k + q_l)]
-\sin[n(q_j + q_k + q_l)]\big\}\times\\
&(A_j e^{i \tau_0 / 2} + c.c.)
(A_k e^{i \tau_0 / 2} + c.c.)
(A_l e^{i \tau_0 / 2} + c.c.),
\end{split}
\label{eq:nonlinearity_term}
\end{equation}

Dissipation term is expressed as:
\begin{equation}
\begin{split}
\Gamma_2 U_n^{(0)^2} U_n^{(0)'} &=\frac{\Gamma_2}{16}\sum_{j,k,l} 
\sin(n q_j) \sin(n q_k) \sin(n q_l)
\big\{(A_j e^{i \tau_0 / 2} + c.c.)\\
&(A_k e^{i \tau_0 / 2} + c.c.)
(iA_l e^{i \tau_0 / 2} + c.c.)\big\}\\
&= \frac{\Gamma_2}{64} 
\sum_{j,k,l} 
\big\{ 
\sin[n(-q_j + q_k + q_l)]+\sin[n(q_j + q_k - q_l)] \\
&+
\sin[n(q_j - q_k + q_l)]
-\sin[n(q_j + q_k + q_l)]\big\}\times\\
&(A_j e^{i \tau_0 / 2} + c.c.)
(A_k e^{i \tau_0 / 2} + c.c.)
(iA_l e^{i \tau_0 / 2} + c.c.),
\end{split}
\label{dissipative_term}
\end{equation}

And finally the detuning can be evaluated as:
\begin{equation}
    \sigma U_n^{(0)} = 
\frac{\sigma}{2} \sum_m (A_m e^{i \tau_0 / 2} + c.c.) \sin(n q_m)
\label{eq:detuning_term}
\end{equation}

Substituting Eqs.~\eqref{eq:second_derivative}-\eqref{eq:detuning_term} into Eq.~\eqref{eq:first_order_eq}, then multiplying by $\sin(n q_m)$ and applying the orthonormality of the modes, the following slow complex mode envelope is given by:

\begin{equation}
\begin{split}
    -\frac{i}{2} \frac{\partial A_m}{\partial \tau_1}
-& \frac{i \Gamma_1}{4} A_m
- \frac{H}{4} A_m^*
+ D \sin^2 \left( \frac{q_m}{2} \right) A_m\\
-& \left( \frac{3 \beta}{32} + \frac{i \Gamma_2}{64} \right)
\sum_{j,k,l} A_j A_k A_l^* \Delta_{j,k,l,m}
+ \frac{\sigma_p}{2} A_m = 0
\end{split}
\end{equation}

\section{\label{app:B} Explit steady-state equation for two oscillator (N=2)}

Starting from the steady state complex equation for $a_m$ Eq.~\eqref{eq:amp_flow_time_independent}:
\begin{equation}
\begin{split}
\Bigg( \sigma_p &+ 2\frac{d}{\omega_p^2} \sin^2 \!\Big( \frac{q_m}{2} \Big)\\
-& \frac{i \gamma}{2\omega_p} \Bigg) a_m
- \frac{h}{2\omega_p^2} a_m^*\\
-& \left( \frac{3 \alpha}{16\omega_p^2} + \frac{i \eta f_n^2}{64\omega_p} \right)
\sum_{j,k,l} a_j a_k a_l^* \, \Delta^{(1)}_{j,k,l,m}
= 0 .
\label{eq:steady_general}
\end{split}
\end{equation}
For \(N=2\) the modal wavenumbers are
\[
q_1=\frac{\pi}{3},\qquad q_2=\frac{2\pi}{3}.
\] 
\subsection*{Evaluate the nonlinear selection-sum for \(N=2\).}
Using the definition of \(\Delta^{(1)}_{j,k,l,m}\) (Eq.\eqref{eq:delta_fn} in the main text) one finds for \(m=1,2\) the compact reductions
\begin{align}
\sum_{j,k,l=1}^2 a_j a_k a_l^*\,\Delta^{(1)}_{j,k,l,1}
&= 3\,a_1|a_1|^2 + 6\,a_1|a_2|^2 + 3\,a_2^2 a_1^*, \label{eq:Delta_sum_m1}\\[4pt]
\sum_{j,k,l=1}^2 a_j a_k a_l^*\,\Delta^{(1)}_{j,k,l,2}
&= 3\,a_2|a_2|^2 + 6\,a_2|a_1|^2 + 3\,a_1^2 a_2^*. \label{eq:Delta_sum_m2}
\end{align}
(These formulas follow from the nonzero index triples for \(N=2\): for \(m=1\) the nonzero triples are \((1,1,1),(1,2,2),(2,1,2),(2,2,1)\), each contributing with coefficient \(3\); similarly for \(m=2\).)\\

Substituting the values of $q_1$, $q_2$ and the delta functions into Eq.~\eqref{eq:steady_general} for \(m=1\) and \(m=2\).

For \(m=1\):
\begin{equation}
\begin{split}
\Bigg(\sigma_p +& 2\frac{d}{2\omega_p^2} - \frac{i\gamma}{2\omega_p}\Bigg)a_1\\
&- \frac{h}{2\omega_p^2}\,a_1^*\\
&- \Bigg(\frac{3\alpha}{16\omega_p^2} + \frac{i\eta f_n^2}{64\omega_p}\Bigg)
\big(3\,a_1|a_1|^2 + 6\,a_1|a_2|^2 + 3\,a_2^2 a_1^*\big) = 0.
\end{split}
\label{eq:N2_m1}
\end{equation}

For \(m=2\):
\begin{equation}
\begin{split}
\Bigg(&\sigma_p + 2\frac{3d}{2\omega_p^2} - \frac{i\gamma}{2\omega_p}\Bigg)a_2\\
&- \frac{h}{2\omega_p^2}\,a_2^*\\
&- \Bigg(\frac{3\alpha}{16\omega_p^2} + \frac{i\eta f_n^2}{64\omega_p}\Bigg)
\big(3\,a_2|a_2|^2 + 6\,a_2|a_1|^2 + 3\,a_1^2 a_2^*\big) = 0.
\end{split}
\label{eq:N2_m2}
\end{equation}
From here, two single-mode solution branches can be obtained by setting $a_1$ or $a_2$ to zero in the above coupled equation. For $a_1$ it is 

\begin{equation}
    \Big[2(\sigma_p\omega_p^2+d)+\frac{9}{8}\alpha\omega_p^2\abs{a_1}^2\Big]^2+\Big[\gamma\omega_p+\frac{3}{32}\eta\omega_p f_n^2\abs{a_1}^2\Big]^2=h^2
    \label{eq:a_1_single_mode}
\end{equation}
and for $a_2$
\begin{equation}
    \Big[2(\sigma_p\omega_p^2+3d)+\frac{9}{8}\alpha\omega_p^2\abs{a_2}^2\Big]^2+\Big[\gamma\omega_p+\frac{3}{32}\eta\omega_p f_n^2\abs{a_2}^2\Big]^2=h^2
    \label{eq:a_2_single_mode}
\end{equation}
The dependency of the response on the drive strength $g$ comes through the term $\sigma_p$ and $f_n$ defined in the main text.

\end{document}